\documentclass[conference,letterpaper]{IEEEtran}
\usepackage[left = 0.64in, right =0.68in, top = 0.73in, bottom = 1.08in]{geometry}
\IEEEoverridecommandlockouts
\usepackage[utf8]{inputenc} 
\usepackage[T1]{fontenc}
\usepackage{url}
\usepackage{ifthen}
\usepackage{cite}
\usepackage[cmex10]{amsmath} 
\usepackage{amssymb}
\usepackage{graphicx}
\usepackage{wrapfig}
\usepackage{multirow}
\usepackage{array}
\usepackage{makecell}
\usepackage{tikz}
\usepackage{algorithm}
\usepackage{algpseudocode}
\usepackage{graphicx}
\usepackage{blkarray}
\usepackage{comment}
\usepackage[english]{babel}
\usepackage{blkarray}
\newtheorem{thm}{Theorem}
\newtheorem{lem}{Lemma}

\newtheorem{defn}{Definition}

\newtheorem{exmp}{Example}

\newtheorem{rem}{Remark}

\usepackage{mathtools}
\newcommand\myeq{\stackrel{\mathclap{\normalfont\mbox{$*$}}}{=}}

\newif\ifcomment
\commenttrue
\begin{document}
	
	\title{Secretive Hotplug Coded Caching}
	\author{\IEEEauthorblockN{Mallikharjuna Chinnapadamala}
\IEEEauthorblockA{\textit{Dept. of Electrical Communication } \\
\textit{Engineering, Indian Institute of Science, } \\
Bengaluru, India \\
Email: chinnapadama@iisc.ac.in}
\and
\IEEEauthorblockN{Charul Rajput}
\IEEEauthorblockA{\textit{Dept. of Mathematics and Systems } \\
\textit{Analysis, Aalto University,}\\
Finland \\
Email: charul.rajput@aalto.fi}
\and
\IEEEauthorblockN{B. Sundar Rajan}
\IEEEauthorblockA{\textit{Dept. of Electrical Communication } \\
\textit{Engineering, Indian Institute of Science, } \\
Bengaluru, India \\
Email: bsrajan@iisc.ac.in}
}

%\author{Mallikharjuna Chinnapadamala and B. Sundar Rajan \\ Department of Electrical Communication Engineering, Indian Institute of Science, Bengaluru \\
%E-mails: chinnapadama@iisc.sc.in, bsrajan@iisc.ac.in}
	%~\IEEEmembership{Staff,~IEEE,}
	% <-this % stops a space
	%\thanks{This paper was produced by the IEEE Publication Technology Group. They are in Piscataway, NJ.}% <-this % stops a space
	%\thanks{Manuscript received April 19, 2021; revised August 16, 2021.}}
	
	
	
	\maketitle
\begin{abstract}

In this work, we consider a coded caching model called \textit{hotplug coded caching}, in which some users are offline during the delivery phase. The concept of Hotplug Placement Delivery Arrays (HpPDAs) for hotplug coded caching systems has been introduced in the literature, and two classes of HpPDAs are known. In this paper, we consider a secrecy constraint in hotplug coded caching setup, where users should not learn anything about any file from their cache content, and active users should not gain any information about files other than their demanded file from either their cache content or the server transmissions. We propose two secretive schemes for the two classes of HpPDAs and compare them with a baseline scheme, which is a secretive scheme using PDAs for the classical coded caching setup and can be trivially adapted for the hotplug coded caching setup. We numerically show that our schemes outperform the baseline scheme in certain memory regions.
\end{abstract}

\section{Introduction}
Coded caching, introduced in \cite{MAN}, leverages content placement in user caches to reduce the communication load during content delivery. The process consists of two phases: (1) the placement phase, which takes place during off-peak hours when the server populates user caches without prior knowledge of their future demands, and (2) the delivery phase, during which user demands are revealed and the server transmits coded multicast messages that simultaneously satisfy multiple users. The server waits till it gets the demands of all the users. We refer to this setup as a classical coded caching setup. The scheme proposed in \cite{MAN}, commonly referred to as the MAN (Maddah-Ali and Niesen) scheme, is proven to be optimal under the constraint of uncoded placement when the number of files is at least equal to the number of users, as shown in \cite{WTP}. When the number of files is smaller than the number of users, an improved scheme is proposed in \cite{YMA}, which achieves better performance by eliminating certain linearly dependent multicast messages that arise when multiple users request the same file.

A major drawback of the MAN scheme is its high subpacketization level, which increases exponentially with the number of users. To address this issue, coded caching schemes with reduced subpacketization levels have been developed using the Placement Delivery Array (PDA) framework introduced in \cite{YCTC}, which integrates placement and delivery into a single array structure. The MAN scheme corresponds to a specific class of PDA known as the MAN-PDA \cite{YCTC}. Additionally, schemes based on linear block codes and block designs, presented in \cite{TaR} and \cite{ASK}, respectively, offer practical subpacketization levels.

The MAN scheme requires that all users submit their demands before the server can begin the delivery phase. This introduces a limitation: it assumes that every user is active and participates synchronously. In scenarios where some users are inactive during the delivery phase, the system must still deliver content to the remaining active users. This challenge is addressed in the hotplug coded caching model introduced in \cite{MaT}. In this model, among $K$ total users, only $K'$ are active during the delivery phase. The server is aware of the number $K'$ during the placement phase, but does not know which specific users will be active.

To address this challenge, \cite{MaT} proposed new coded caching schemes based on Maximum Distance Separable (MDS) coded placement, referred to as the MT (Ma and Tuninetti) scheme. Hotplug coded caching under the constraint of demand privacy, ensuring that no user can infer the demand of any other user, was studied in \cite{MaT2}.
 Additionally, the notion of Placement Delivery Arrays (PDAs) was extended to hotplug coded caching in \cite{RaR}, which introduced Hotplug PDAs (HpPDAs). The same authors presented a construction method for HpPDAs using combinatorial designs known as $t$-designs in \cite{RaR2}.

In the dedicated cache setup presented in \cite{MAN}, an additional constraint was introduced in \cite{RPKP} to ensure that users cannot obtain any information about files other than their requested ones, either from their local caches or from the transmissions sent by the server. This concept is referred to as private coded caching in \cite{RPKP} and as secretive coded caching in \cite{RPKP2}. we adopt the latter terminology throughout this work. The notion of secretive coded caching has since been extended to various network configurations, including shared cache networks \cite{MeR}, \cite{PNR}, device-to-device networks \cite{ZY}, combination networks \cite{ZY2}, and fog radio access networks \cite{TJHZN}, multi-access networks \cite{ChR}. Moreover, the problem of secretive coded caching in dedicated cache setups with colluding users has been studied in \cite{MaS}. To the best of our knowledge, secrecy has not been explored in the hotplug coded caching system.
\subsection{Our contribution}
We consider secrecy in the hotplug coded caching system and propose two schemes for two classes of HpPDAs: (1) MAN-HpPDAs, and (2) HpPDAs from $t$-designs.
We compare the performance of our schemes with that of the baseline scheme, which is the secretive scheme using PDAs for the classical coded caching setup. We numerically show that our schemes perform better than the baseline scheme in certain memory regions.

This paper is organised as follows: Section \ref{sec2} introduces the problem setup. Some preliminaries are presented in Section \ref{sec3}. Section \ref{sec4} provides the main results. The proofs of Theorem \ref{thm1} and Theorem  \ref{thm2} are given in Section \ref{scheme1} and Section \ref{scheme2} respectively. Numerical comparisons are given in Section \ref{num}, and Section \ref{conc} concludes the paper.

\subsection{Notations}
For a positive integer $n$, $[n]$ denotes the set $\{1,2,\cdots,n\}$. For two sets $A$ and $B$, the notation $A \setminus B$ denotes the set of elements in $A$ which are not in $B$. For a set $A$, the number of elements in it is represented by $|A|$. The binomial coefficient  $\binom{n}{k}$ is equal to $\frac{n!}{k!(n-k)!}$, where $n$ and $k$ are positive integers such that $k \leq n$. For a set $S$ and a positive integer $t$, the notation ${S \choose t}$ represents the collection of all the subsets of $S$ of size $t$. For a matrix $\textbf{D}$, $[\textbf{D}]_{ij}$ represents the $(i,j)^{th}$ entry of $\textbf{D}$.

\section{Problem Setup} \label{sec2}

In a $(K, K', N)$ hotplug coded caching system, the server stores $N$ files, denoted by $W_{[N]}=W_{1},W_{2},\ldots,W_{N}$, each of size $B$ bits. It is connected to $K$ users through an error-free shared link. Each user is equipped with a cache of size $M$ files. $Z_{k}$ denotes the content of the cache of user $k \in [K]$. The system has a placement phase and a delivery phase.

\textit{Placement Phase}: We assume that the server knows that $K'\leq K$ users will be active during the delivery phase. During the placement phase, the server is unaware of which $K'$ users will be active. The server fills the cache as a function of the file library $W_{[N]}$ and a randomness $V$ of appropriate size, such that
\begin{equation} \label{eqsec1}
I(W_{[N]}; Z_{k}) = 0 , \quad \forall~ k \in [K].
\end{equation}
Equation (\ref{eqsec1}) implies that no user knows anything about any file from the content of its cache. The placement is carried out without knowing the future demands of the users.

\textit{Delivery Phase}: During this phase, the set of active users, denoted by $\mathcal{I} \in \binom{[K]}{K'}$, sends their demands to the server. The demand of user $k$ is denoted by $d_k$. 

Upon knowing the set of active users and their demands, the server transmits a message $X$ over the shared link. Each active user $k \in \mathcal{I}$ must be able to reconstruct their requested file $W_{d_k}$ using the transmission $X$ and their cache content $Z_k$, while learning nothing about the remaining $N-1$ files. That is,
\begin{equation} \label{eqdec}
H(W_{d_k} \mid X, Z_k) = 0, \quad \forall\, k \in \mathcal{I},
\end{equation}
\begin{equation} \label{eqsec2}
I(W_{[N] \setminus \{d_k\}};\, X, Z_k) = 0, \quad \forall\, k \in \mathcal{I}.
\end{equation}

The total size of files sent by the server is called the rate, denoted by $R$, of the system.

\begin{defn} \label{def1}
We say that the pair $(M, R)$ is secretively achievable if there exists a scheme that satisfies the conditions in (\ref{eqsec1}), (\ref{eqdec}), and (\ref{eqsec2}) with rate $R$ and memory $M$. We refer to the scheme as a secretive hotplug coded caching scheme. The optimal rate for the given setting is defined as:
\begin{equation} \label{eq9}
R^{*} = \inf\{R : (M, R) \text{ is achievable} \}.
\end{equation}
\end{defn}

Since a $(K,K',N)$ hotplug system cannot perform better than a classical coded caching system with $K'$ users, any converse bound from the classical coded caching system with $K'$ users and $N$ files is applicable to the $(K,K',N)$ hotplug system.

The following lower bound was given in \cite{RPKP}.

\begin{lem}[\cite{RPKP}] \label{bound}
For $N$ files and $K'$ users, each with cache size $1\leq M\leq N(K'-1)$,
\begin{equation}
R^{*} \geq \max_{l \in \{1,2,\ldots,\min(N/2,K')\}} \frac{l \lfloor N/l \rfloor - 1 - (l-1) M}{\lfloor N/l \rfloor - 1}.
\end{equation}
\end{lem}

By substituting $l=1$ in the above inequality, we get $R^{*}\geq 1$. So, this is the minimum achievable rate to satisfy the secrecy condition. This implies that the user cannot know anything about the requested file from the cached content alone and needs $B$ bits from the server to learn anything about it.
\section{Preliminaries} \label{sec3}
In this section, we discuss some preliminaries. We first review the definitions of PDAs for the classical coded caching system and the hotplug coded caching system. Then, we present the definitions of a design and a $t$-design. Next, we explain the idea of non-perfect secret sharing and define Cauchy matrices. Finally, we briefly describe the secretive coded caching scheme based on PDAs, as given in \cite{MeR2}.

\subsection{Placement Delivery Arrays}
\begin{defn}[Placement Delivery Array \cite{YCTC}]
	For positive integers $K, F, Z$ and $S$, an $F \times K$ array $\mathbf{P} = (p_{j,k})_{j \in[F], k \in [K]}$, composed of a special symbol ``$*$" and $S$ non-negative integers from $[S]$, is called a $[K, F, Z, S]$-PDA if it satisfies the following conditions:
	\begin{enumerate}
		\item The symbol ``$*$" appears $Z$ times in each column,
		\item Each integer occurs at least once in the array,
		\item For any two distinct entries $p_{j_1,k_1}$ and $p_{j_2, k_2}$, $p_{j_1, k_1} =
		p_{j_2, k_2} = s$ is an integer only if
		\begin{enumerate}
			\item $j_1 \neq j_2,~ k_1 \neq k_2$, i.e., they lie in distinct rows and distinct columns; and
			\item $p_{j_1, k_2} = p_{j_2, k_1} = *$, i.e., the corresponding $2 \times 2$ sub-array formed by rows $j_1, j_2$ and columns $k_1, k_2$ must be of the following form:
			\[
			\begin{bmatrix}
			s & * \\
			* & s
			\end{bmatrix} \quad
			\text{or} \quad
			\begin{bmatrix}
			* & s \\
			s & *
			\end{bmatrix}.
			\]
		\end{enumerate}
	\end{enumerate}
\end{defn}

A $[K, F, Z, S]$-PDA $\mathbf{P}$ corresponds to a $(K, N, M)$ coded caching scheme with $K$ users, $N$ files, cache memory of size $M$ files, subpacketization level $F$, $\frac{M}{N}=\frac{Z}{F}$ and rate $R=\frac{S}{F}$.

\begin{rem}
	A PDA $\mathbf{P}$ is called a $g$-$[K, F, Z, S]$ regular PDA if each integer $s\in [S]$ appears exactly $g$ times in $\mathbf{P}$. The MAN scheme corresponds to a special class of PDAs called MAN-PDAs.
\end{rem}

\begin{defn}[Hotplug Placement Delivery Array (HpPDA)\cite{RaR}]
	Let $K, K', F, F', Z, Z'$ and $S$ be integers such that $K \geq K'$, $F \geq F'$, and $Z < F'$. Consider two arrays given as follows:
	\begin{itemize}
		\item $\mathbf{P}=(p_{f,k})_{f \in [F], k \in [K]}$ is an array containing ``$*$" and null entries. Each column contains $Z$ occurrences of ``$*$",
		\item $\mathbf{B}=(b_{f,k})_{f \in [F'], k \in [K']}$ is a $[K', F', Z', S]$-PDA.
	\end{itemize}
	
	For each $\tau \subseteq [K]$, with $|\tau| = K'$, there exists a subset $\zeta \subseteq [F]$, with $|\zeta|=F'$ such that
	\begin{equation}\label{c}
		[\mathbf{P}]_{\zeta \times \tau} \myeq \mathbf{B},
	\end{equation}
	where $[\mathbf{P}]_{\zeta \times \tau}$ denotes the subarray of $\mathbf{P}$ whose rows correspond to the set $\zeta$ and columns correspond to the set $\tau$, and $\myeq$ denotes that the positions of all ``$*$" entries are the same in both arrays. We call $(\mathbf{P}, \mathbf{B})$ a $(K,K',F,F',Z,Z',S)$-HpPDA.
\end{defn}

Two classes of HpPDAs were introduced in \cite{RaR2}: (1) MAN-HpPDAs, and (2) HpPDAs constructed from $t$-designs. The definition of a design is given below.
\subsection{ Combinatorial designs \cite{Sti}}
\begin{defn}[Design]
A design is a pair $(X, \mathcal{A})$ such that the following properties are satisfied:
\begin{enumerate}
\item $X$ is a set of elements called points, and
\item $\mathcal{A}$ is a collection of nonempty subsets of $X$ called blocks.
\end{enumerate}
\end{defn}
\begin{defn}[$t$-design]
Let $v, k, \lambda$, and $t$ be positive integers such that $v > k \geq t$. A $t$-$(v, k, \lambda)$ design is a design $(X, \mathcal{A})$ such that:
\begin{enumerate}
\item $|X| = v$,
\item each block contains exactly $k$ points, and
\item every set of $t$ distinct points is contained in exactly $\lambda$ blocks.
\end{enumerate}
\end{defn}

Let $(X,\mathcal{A})$ be a $t$-$(v,k,\lambda)$ design with non-repeated blocks. From \cite{Sti}, it follows that, for  $Y \subseteq X$, where $|Y| = s \leq t$, there are exactly $\lambda_s =\frac{\lambda{v-s \choose t-s}}{{k-s \choose t-s}}$ blocks in $\mathcal{A}$ that contain all the points in $Y$. For $ Y \subseteq T \subseteq X$, where $|T| = t, |Y| = i$ with $i \leq t$, there are exactly
$\lambda_i^{t} =\frac{\lambda{v-t \choose k-i}}{{v-t \choose k-t}}$
blocks in $\mathcal{A}$ that contain all the points in $Y$ and none of the points in $T \backslash Y$.
\subsection{Secret Sharing Schemes}
Coded caching schemes that study secrecy in the literature use non-perfect secret sharing schemes \cite{CDN}. The key idea is to encode the secret into shares in such a way that accessing a certain number of shares does not reveal any information about the secret, while accessing all the shares enables complete recovery. The formal definition of a non-perfect secret-sharing scheme is given below.

\begin{defn}[\cite{CDN}]
For a secret $W$ of size $B$ bits and $m < n$, an $(m,n)$ non-perfect secret sharing scheme generates $n$ equal-sized shares $S_{1},S_{2},\ldots,S_{n}$ such that:
\begin{subequations}
\begin{align} \label{eq10}
&I(W;\mathbf{S}) = 0, \quad \forall  \ \mathbf{S} \subseteq \{S_{1},S_{2},\ldots,S_{n} \}, ~\text{such that}~ |\mathbf{S}| \leq m, \\
&H(W|S_{1},S_{2},\ldots,S_{n}) = 0.
\end{align}
\end{subequations}
\end{defn}

The size of each share should be at least $\frac{B}{n-m}$ bits in a non-perfect secret sharing scheme \cite{RPKP}. For large enough $B$, there exist non-perfect secret sharing schemes with share size exactly equal to $\frac{B}{n-m}$ bits.

\subsection{Cauchy Matrix}

\begin{defn}[\cite{PlX}]
Let $\mathcal{F}_{q}$ be a finite field with $q \geq u+v$. Let $X = \{x_{1}, x_{2}, \ldots, x_{u} \}$ and $Y = \{y_{1}, y_{2}, \ldots, y_{v} \}$ be two disjoint subsets of $\mathcal{F}_{q}$ such that:
\begin{itemize}
\item For all $i \in [u], j \in [v]: x_i - y_j \neq 0$,
\item For all $i \neq j \in [u]: x_i \neq x_j$, and all $i \neq j \in [v]: y_i \neq y_j$.
\end{itemize}
Then, the matrix
\[
A =
\begin{bmatrix}
\frac{1}{x_1 - y_1} & \frac{1}{x_1 - y_2} & \cdots & \frac{1}{x_1 - y_v} \\
\frac{1}{x_2 - y_1} & \frac{1}{x_2 - y_2} & \cdots & \frac{1}{x_2 - y_v} \\
\vdots & \vdots & \ddots & \vdots \\
\frac{1}{x_u - y_1} & \frac{1}{x_u - y_2} & \cdots & \frac{1}{x_u - y_v}
\end{bmatrix}
\]
is a Cauchy matrix of dimension $u \times v$. A Cauchy matrix has full rank, and each of its submatrices is also a Cauchy matrix.
\end{defn}

\subsection{Secretive Coded Caching Scheme from PDAs \cite{MeR2}}

A secretive coded caching scheme for classical coded caching systems using PDAs was proposed in \cite{MeR2}. For any given $(K, F, Z, S)$-PDA, the achievable scheme is as follows: Each file is encoded using a $(Z,F)$ non-perfect secret sharing scheme \cite{CDN}. The server places the shares, along with randomly generated keys (each of the same length as a share), in each user's cache during the placement phase. Once all users reveal their demands to the server, it sends $S$ transmissions over the shared link. Each transmission is an XOR of shares, further XOR-ed with a key. For $M=\frac{NZ}{F-Z}+1$, the secretively achievable rate $R(M)$ is given by:
\[
R(M) = \frac{S}{F-Z}.
\]
 We refer to this scheme as the baseline scheme. We use the baseline scheme to compare the performance of the hotplug coded caching schemes.

 \section{Main Results} \label{sec4}
 In this section, we present the main results of this work. 
 \begin{thm} \label{thm1}
    For the $(K, K', N)$ hotplug coded caching system, there exists a secretive hotplug coded caching scheme using a $(K, K', F, F', Z, Z', S)$-MAN-HpPDA with the memory-rate pair given by 
    \[
    (M,R) = \left(\frac{NZ+\binom{K-1}{t}}{F'-Z'}, \frac{S}{F'-Z'}\right),
    \]
    where $t \in [0: K'-2]$. For $M = N \binom{K-1}{K'-2}$, the rate $R = 1$ is achievable.
\end{thm}
\textit{Proof}: The detailed scheme that proves the above theorem is given in Section \ref{scheme1}.

This scheme outperforms the baseline scheme in the low memory region. A numerical comparison is provided in Section~\ref{num}.
\begin{thm}\label{thm2}
For the $(K, K', N)$ hotplug coded caching system, there exists a secretive hotplug coded caching scheme using a $(K, K', F, F', Z, Z', S)$-HpPDA designed from a $t$-$(v,k,\lambda)$ design with the following memory-rate pair
\[
(M,R)=\left (\frac{NZ}{F'-Z'}+\frac{1}{F'-Z'} \sum_{s=1}^{t-2} a_s \binom{K-1}{s}, \frac{S}{F'-Z'}\right),
\]
 where $0 \leq a_s \leq \lambda_s^{t}$, $1 \leq s \leq t-1.$
\end{thm}
\textit{Proof}: The detailed scheme that proves the above theorem is given in Section \ref{scheme2}.

A numerical comparison of the scheme is presented in Section~\ref{num}, where we demonstrate that the scheme in Theorem~\ref{thm2} outperforms the baseline scheme in a certain memory region.
\section{Proof of Theorem \ref{thm1}}\label{scheme1}
In this section, we prove Theorem \ref{thm1}. We present a secretive hotplug coded caching scheme using MAN-HpPDAs.

Before presenting the scheme, we briefly discuss MAN-HpPDAs. For integers $K$, $K'$, and $t$ such that $K' < K$ and $t \in [K']$, there exists a $(K, K', F, F', Z, Z', S)$ MAN-HpPDA $(\mathbf{P}, \mathbf{B})$, where 
\begin{align*}
F' &= \binom{K'}{t}, \quad F = \binom{K}{t}, \quad Z' = \binom{K'-1}{t-1}, \\
Z &= \binom{K-1}{t-1}, \quad S = \binom{K'}{t+1}.
\end{align*}

Here, the array $B$ is a $(K', F', Z', S)$ MAN-PDA for $K'$ users. To construct the array $\mathbf{P}$, consider an MAN-PDA with parameters $(K, F, Z, S')$ for $K$ users and replace all the integers with null symbols. Let us index the rows of $\mathbf{P}$ using all $t$-sized subsets of $[K]$, arranged in lexicographic order.
%Define the set of all $t$-sized subsets of $[K]$ as
%\begin{equation} \label{eq1}
 %   \Omega_{t} \triangleq \{\mathcal{G} : \mathcal{G} \subseteq [K], |\mathcal{G}| = t\}.
%\end{equation}
The Placement and delivery phases are described below.

\textit{Placement Phase}: For $t\in[K'-2]$, each file is encoded using a $\left(\binom{K-1}{t-1},\binom{K'-1}{t}+\binom{K-1}{t-1}\right)$ non-perfect secret sharing scheme \cite{CDN}. To do this, each file is divided into $\binom{K'-1}{t}$ equal-sized parts. Let $m=\binom{K-1}{t-1}$, and $n=\binom{K'-1}{t}+\binom{K-1}{t-1}$. So,
 \begin{equation} \label{eq18}
    W_{i}=\{W_{i,h}: h\in [n-m]\}, \forall~ i \in [N].
\end{equation}
Now generate $m$ encryption keys for each $i\in [N]$ independently and uniformly from $\mathcal{F}_{2}^{B/\binom{K'-1}{t}}$. Let those keys be $\{Y_{i,p}: i\in[N], p\in [m]\}$. Consider a $n\times n$ cauchy matrix\cite{PlX}, $\textbf{A}$. Let $[\textbf{A}]_{u,v}=a_{u,v}, \ \forall~ u,v \in [n]$ over $\mathcal{F}_{2^{l}}$, where $2^{l} \geq 2n $. The sub-matrices of a Cauchy matrix have
full rank. For each file $W_{i}, i \in [N]$, the shares $\{\tilde{W}_{i,s}: s \in [n] \}$ are generated as follows:
%\[
\begin{equation}\label{eqshares}
\begin{bmatrix}
    \tilde{W}_{i,1} \\
    \tilde{W}_{i,2} \\
    \vdots\\
    \tilde{W}_{i,n}
\end{bmatrix} 
=
\begin{bmatrix}
    a_{1,1} &a_{1,2} &\cdots &a_{1,n} \\
    a_{2,1} &a_{2,2} &\cdots &a_{2,n} \\
    \vdots &\vdots &\ddots &\vdots \\
    
    a_{n,1} &a_{n,2} &\cdots & a_{n,n}
    
\end{bmatrix}
\cdot
\begin{bmatrix}
    W_{i,1} \\
    . \\
    W_{i,n-m} \\
    Y_{i,1} \\
    . \\
    Y{i,m}
\end{bmatrix}
\end{equation}
%\]
Now, consider a $\left(\binom{K}{t}, \binom{K'-1}{t}+\binom{K-1}{t-1}\right)$ Maximum Distance Separable (MDS) code. For each file, using the $n$ shares, $\binom{K}{t}$ coded shares are generated through this MDS code. Let the coded shares be denoted by $\{C_{n,\mathcal{T}}: \mathcal{T} \subset [K], |\mathcal{T}|=t\}$. The size of each coded share is equal to that of a single share.

Next, generate $\binom{K}{t+1}$ random vectors independently and uniformly from $\mathbb{F}_{2}^{B/\binom{K'-1}{t}}$. Let these vectors be denoted by $\{V_{\mathcal{S}}: \mathcal{S} \subset [K], |\mathcal{S}|=t+1\}$.

The cache placement for cache $k \in [K]$ is done as follows:
\begin{multline}
Z_{k} = \{C_{i,\mathcal{T}} : \mathbf{P}_{\mathcal{T},k}=*, i \in [N], \mathcal{T} \in \Omega_t \} ~\cup \\ \{V_{\mathcal{S}} : k \in \mathcal{S},~ \mathcal{S} \subset [K],~ |\mathcal{S}| = t+1\}.
\end{multline}
Equivalently, we can write the cache content as follows:
\begin{multline}
Z_{k} = \{C_{i,\mathcal{T}} : k \in \mathcal{T},~ |\mathcal{T}| = t, i \in [N]\} ~\cup \\ \{V_{\mathcal{S}} : k \in \mathcal{S},~\mathcal{S} \subset [K],~ |\mathcal{S}| = t+1\}.
\end{multline}

According to this placement, each cache stores $\binom{K-1}{t-1}$ coded shares of every file, with each coded share having size $\frac{1}{\binom{K'-1}{t}}$ (in units of a file). Each cache also stores $\binom{K-1}{t}$ random vectors, each of size $\frac{1}{\binom{K'-1}{t}}$. Therefore, the total size of each cache is
\begin{equation}
M = \frac{N\binom{K-1}{t-1}}{\binom{K'-1}{t}} + \frac{\binom{K-1}{t}}{\binom{K'-1}{t}}.
\end{equation}

\textit{Delivery Phase}: Let $\mathcal{I} = \{i_1, i_2, \ldots, i_{K'}\}$, where $1 \leq i_k \leq K$ for all $k \in [K']$, denote the set of active users. The demand vector is given by $\textbf{d} = (d_{i_1}, d_{i_2}, \ldots, d_{i_{K'}})$. By the property of $(\mathbf{P}, \mathbf{B})$ HpPDA, for $\mathcal{I} \subseteq [K], |\mathcal{I}| = K'$, there exists a subset $\zeta \subseteq \binom{[K]}{t}, |\zeta|=\binom{K'}{t}$ such that $[\mathbf{P}]_{\zeta \times \mathcal{I}} \myeq \mathbf{B}$. Make a new array $\overline{\mathbf{P}}=(\overline{p}_{\mathcal{T},k})_{\mathcal{T} \in \zeta, k \in \mathcal{I}}$ by filling $s \in [S]$ integers in null spaces of the subarray $[\mathbf{P}]_{\zeta \times \mathcal{I}}$ in such a way that $\overline{\mathbf{P}}=\mathbf{B}$. The server sends a transmission for every $s \in [S]$. As $S=\binom{K'}{t+1}$ in MAN-HpPDA,  the server transmits, for every subset $\mathcal{S} \subset \mathcal{I}$ with $|\mathcal{S}| = t+1$, the following message:
\begin{equation} \label{eq27}
X_{\mathcal{S}} = V_{\mathcal{S}} \oplus \bigoplus_{\substack{k \in \mathcal{S}}} C_{d_k,\, \mathcal{S} \setminus \{k\}}.
\end{equation}
 The server sends a transmission for every $s \in [S]$. Each transmission is of size $\frac{1}{F'-Z'}$. So, the achieved rate is $\frac{S}{F'-Z'}$.

\textit{Correctness}: From the transmissions sent by the server, the active users must be able to get their demanded files. Consider a user $k\in \mathcal{I}$ and the transmission $X_{\mathcal{S}}$ such that $k\in \mathcal{S}$. The user $k$ has access to the coded shares $C_{i,\mathcal{S}\backslash h}, h \in \mathcal{S}, h \neq k$ for all $i\in [N]$ from its cache. It also has access to the random vector $V_{\mathcal{S}}$ where $k\in \mathcal{S}$ from its cache. So, the user $k$ gets the coded share $C_{{d}_{k},\mathcal{S} \backslash k}$. From the transmissions $\{ X_{\mathcal{S}} : k \in \mathcal{S}\}$, the user $k$ gets $\binom{K'-1}{t}$ coded shares of its demanded file. There are $\binom{K-1}{t-1}$ coded shares available from its cache. So, it has $\binom{K'-1}{t}+\binom{K-1}{t-1}$ coded shares of its demanded file. As we are using $(\binom{K}{t}, \binom{K'-1}{t}+\binom{K-1}{t-1})$ MDS code, with $\binom{K'-1}{t}+\binom{K-1}{t-1}$ coded shares the user can get $\binom{K'-1}{t}+\binom{K-1}{t-1}$ shares. By the property of a $(\binom{K-1}{t-1}, \binom{K'-1}{t}+\binom{K-1}{t-1})$ non-perfect secret sharing scheme \cite{CDN}, the user $k$ will be able to get back its demanded file. Similarly, all the other active users can get back their demanded files.\\
\textit{Proof of Secrecy}: No user must gain any information about any file using only the contents of the cache that is accessible to it. By using the cache contents and the signals transmitted by the server, no active user should gain any information about files it did not request.
To prove secrecy, first, let's show that a user cannot gain any information about any file using its cache contents alone. We make use of the procedure given in \cite{MaS} to prove this. The shares of all files are generated in the same way. let $l_{f} \in [n], f \in [m]$. Consider any $m$ shares of the file $W_{i}, i\in [N]$. They were generated as follows:  
\[
\begin{bmatrix}
    \tilde{W}_{i,l_{1}} \\
    \tilde{W}_{i,l_{2}} \\
    \vdots \\
    \tilde{W}_{i,l_{m}}
\end{bmatrix} 
=
\begin{bmatrix}
    a_{l_{1},1} &a_{l_{1},2} &\cdots &a_{l_{1},n} \\
    a_{l_{2},1} &a_{l_{2},2} &\cdots &a_{l_{2},n} \\
    \vdots &\vdots &\ddots &\vdots \\
    
    a_{l_{m},1} &a_{l_{m},2} &\cdots & a_{l_{m},n}
    
\end{bmatrix}
\cdot
\begin{bmatrix}
    W_{i,1} \\
    . \\
    W_{i,\binom{K'-1}{t}} \\
    Y_{i,1} \\
    . \\
    Y{i,m}
\end{bmatrix}
\]
Let $\textbf{A}_{1}$ and $\textbf{A}_{2}$ be two sub-matrices of $\textbf{A}$ of size $m\times (n-m)$ and $m\times m$ respectively. Then, the above equation can be written as 
\[
\begin{bmatrix}
    \tilde{W}_{i,l_{1}} \\
    \tilde{W}_{i,l_{2}} \\
    \vdots \\
    \tilde{W}_{i,l_{m}}
\end{bmatrix} 
=
\textbf{A}_{1}   \begin{bmatrix}
    W_{i,1} \\
    W_{i,2} \\
    \vdots \\
    W_{i,\binom{K'-1}{t}} \\
    
\end{bmatrix} + \textbf{A}_{2}
\begin{bmatrix}
    Y_{i,1} \\
    Y_{i,1} \\
    \vdots \\
    Y{i,m}
\end{bmatrix}
\]

For the collection to leak the information, there should be a non-zero matrix $\textbf{H}$ such that
\begin{equation*}
    \textbf{H}\textbf{A}_{1} \neq \textbf{0} ,\hspace{1cm}  \textbf{H}\textbf{A}_{2} = \textbf{0},   
\end{equation*}
where \textbf{0} is the zero matrix. However, the submatrix of a Cauchy matrix is a full-rank matrix. The rows of $\textbf{A}_{2}$ are linearly independent, which implies the non-existence of such \textbf{H}. Thus, information cannot be leaked from the shares of files available to each user from its cache. Now, consider the transmissions sent by the server. Each transmission is protected by a key that is available only to the user for which that particular transmission is useful.  Thus, the scheme achieves secrecy. 
\subsection{Scheme when $t=K'-1$}
When $t = K'-1$, the server makes only one transmission, which means that all active users require that transmission. In this case, there is no need to protect the transmission from any user. To encode the file, a $( \binom{K-1}{K'-2},\, 1 + \binom{K-1}{K'-2} )$ non-perfect secret sharing scheme~\cite{CDN} is used. Then, the server generates $\binom{K}{K'-1}$ coded shares using a $( \binom{K}{K'-1},\, 1 + \binom{K-1}{K'-2} )$ MDS (Maximum Distance Separable) code.
The cache placement for each user $k \in [K]$ is defined as follows:
\begin{equation} \label{eq30}
Z_{k} = \{\tilde{W}_{i,\mathcal{T}} : k\in\mathcal{T} , \mathcal{T} \in \binom{[K]}{K'-1}  \quad \forall \quad i \in [N] \}.
\end{equation}
By the above placement, the size of each cache is
\begin{equation} \label{eq31}
    M= N\binom{K-1}{K'-2}.
\end{equation}
Let the set of active users be denoted by $\mathcal{I}$. During the delivery phase, the server transmits
\begin{equation*}
X = \bigoplus_{\substack{k \in \mathcal{I}}} C_{d_k, [K'] \setminus \{k\}}.
\end{equation*}

Each active user requires only one coded share. It is clear that every user can recover the required coded share from the server's transmission, as it has access to all other coded shares except the one needed. 

The secrecy condition is also satisfied, since each user has $\binom{K-1}{K'-2}$ coded shares in its cache. By the nature of the $\left(\binom{K-1}{K'-2}, 1+\binom{K-1}{K'-2}\right)$ non-perfect secret sharing scheme~\cite{CDN}, a user gains no information about the file from only $\binom{K-1}{K'-2}$ coded shares. The achieved rate is $1$, which is optimal. This completes the proof of Theorem \ref{thm1}. \hfill $\blacksquare$

\begin{exmp}
 Consider a $(6,4,15,6,5,3,3)$-HpPDA corresponding to a hotplug coded caching system with $K = 6$ users, $N=6$ files, and $K' = 4$ active users. For $t = 2$, we have $F = 15$, $F' = 6$, $Z = 5$, $Z' = 3$, and $S = 3$. The corresponding HpPDA $(\mathbf{P}, \mathbf{B})$ is given below. 
\begin{align*}
\mathbf{B} &= \begin{bmatrix}
* & * & 1 & 2 \\
* & 1 & * & 3 \\
* & 2 & 3 & * \\
1 & * & * & 4 \\
2 & * & 4 & * \\
3 & 4 & * & *
\end{bmatrix}, \quad
\mathbf{P} = 
\begin{matrix}
\{1,2\} \\ \{1,3\} \\ \{1,4\} \\ \{1,5\} \\ \{1,6\} \\ \{2,3\} \\ \{2,4\} \\ \{2,5\} \\ \{2,6\} \\ \{3,4\} \\ \{3,5\} \\ \{3,6\} \\ \{4,5\} \\ \{4,6\} \\ \{5,6\}
\end{matrix}
\begin{bmatrix}
* & * &   &   &   &   \\
* &   & * &   &   &   \\
* &   &   & * &   &   \\
* &   &   &   & * &   \\
* &   &   &   &   & * \\
  & * & * &   &   &   \\
  & * &   & * &   &   \\
  & * &   &   & * &   \\
  & * &   &   &   & * \\
  &   & * & * &   &   \\
  &   & * &   & * &   \\
  &   & * &   &   & * \\
  &   &   & * & * &   \\
  &   &   & * &   & * \\
  &   &   &   & * & *
\end{bmatrix}.
\end{align*}
Clearly, for each $\tau \subseteq [6]$ with $|\tau| = 4$, there exists a set $\zeta = \{ S \subseteq \tau \mid |S| = 2 \}$ of size $|\zeta| = 6$ such that the subarray $[\mathbf{P}]_{\zeta \times \tau}$ is equivalent to $\mathbf{B}$, i.e., $[\mathbf{P}]_{\zeta \times \tau} \myeq \mathbf{B}$. Here, the rows of $\mathbf{P}$ are indexed by all 2-element subsets of $[6]$, and the columns are indexed by the elements of $[6]$.

\textit{Placement Phase}: Each file is divided into $3$ parts and then encoded using a $(5,8)$ non-perfect secret sharing scheme. As a result, each file has $8$ shares. Then, using these $8$ shares, $15$ coded shares are generated through a $(15,8)$ MDS code. The $15$ coded shares of file $W_{i}, i\in [N]$ are $\{ C_{i,\{1,2\}}, C_{i,\{1,3\}}, C_{i,\{1,4\}}, C_{i,\{1,5\}}, C_{i,\{1,6\}}, C_{i,\{2,3\}}, C_{i,\{2,4\}},\\ C_{i,\{2,5\}}, C_{i,\{2,6\}}, C_{i,\{3,4\}}, C_{i,\{3,5\}}, C_{i,\{3,6\}}, C_{i,\{4,5\}}, C_{i,\{4,6\}},\\ C_{i,\{5,6\}} \}$. Then, generate $\binom{6}{3}=20$ random vectors uniformly and independently from $\mathcal{F}_{2}^{B/3}$. Let those vectors be represnted as $\{ V_{\mathcal{S}}: \mathcal{S} \in [6], |\mathcal{S}|=3\}$. The placement $\forall i \in [6]$ is done as follows
\begin{align*}
Z_{1}= &\{ C_{i,\{1,2\}}, C_{i,\{1,3\}}, C_{i,\{1,4\}}, C_{i,\{1,5\}}, C_{i,\{1,6\}} \} ~\cup \\ & \{ V_{\{1,2,3\}}, V_{\{1,2,4\}}, V_{\{1,2,5\}}, V_{\{1,2,6\}},  V_{\{1,3,4\}}, V_{\{1,3,5\}},\\ & V_{\{1,3,6\}},  V_{\{1,4,5\}},  V_{\{1,4,6\}}, V_{\{1,5,6\}}\}.\\ 
%\end{align*}
%\begin{align*}
Z_{2}= &\{ C_{i,\{1,2\}}, C_{i,\{2,3\}}, C_{i,\{2,4\}}, C_{i,\{2,5\}}, C_{i,\{2,6\}} \} ~\cup \\ & \{ V_{\{1,2,3\}}, V_{\{1,2,4\}}, V_{\{1,2,5\}}, V_{\{1,2,6\}},  V_{\{2,3,4\}}, V_{\{2,3,5\}},\\ & V_{\{2,3,6\}},  V_{\{2,4,5\}},  V_{\{2,4,6\}}, V_{\{2,5,6\}}\}. \\
Z_{3}= &\{ C_{i,\{1,3\}}, C_{i,\{2,3\}}, C_{i,\{3,4\}}, C_{i,\{3,5\}}, C_{i,\{3,6\}} \} ~\cup \\ & \{ V_{\{1,2,3\}}, V_{\{1,3,4\}}, V_{\{1,3,5\}}, V_{\{1,3,6\}},  V_{\{2,3,4\}}, V_{\{2,3,5\}},\\ & V_{\{2,3,6\}},  V_{\{3,4,5\}},  V_{\{3,4,6\}}, V_{\{3,5,6\}}\}. \\
Z_{4}= &\{ C_{i,\{1,4\}}, C_{i,\{2,4\}}, C_{i,\{3,4\}}, C_{i,\{4,5\}}, C_{i,\{4,6\}} \} ~\cup \\ & \{ V_{\{1,2,4\}}, V_{\{1,3,4\}}, V_{\{1,4,5\}}, V_{\{1,4,6\}},  V_{\{2,3,4\}}, V_{\{2,4,5\}}, \\ & V_{\{2,4,6\}},  V_{\{3,4,5\}},  V_{\{3,4,6\}}, V_{\{4,5,6\}}\}. \\
Z_{5}= &\{ C_{i,\{1,5\}}, C_{i,\{2,5\}}, C_{i,\{3,5\}}, C_{i,\{4,5\}}, C_{i,\{5,6\}} \} ~\cup \\ & \{ V_{\{1,2,5\}}, V_{\{1,3,5\}}, V_{\{1,4,5\}}, V_{\{1,5,6\}},  V_{\{2,3,5\}}, V_{\{2,4,5\}}, \\ & V_{\{2,5,6\}},  V_{\{3,4,5\}},  V_{\{3,5,6\}}, V_{\{4,5,6\}}\}. \\
Z_{6}= &\{ C_{i,\{1,6\}}, C_{i,\{2,6\}}, C_{i,\{3,6\}}, C_{i,\{4,6\}}, C_{i,\{5,6\}} \} ~\cup \\ & \{ V_{\{1,2,6\}}, V_{\{1,3,6\}}, V_{\{1,4,6\}}, V_{\{1,5,6\}},  V_{\{2,3,6\}}, V_{\{2,4,6\}},\\ & V_{\{2,5,6\}},  V_{\{3,4,6\}},  V_{\{3,5,6\}}, V_{\{4,5,6\}}\}.
\end{align*}

By the above placement, the size of each cache is $\frac{40}{3}$.\\
\noindent \textit{Delivery Phase:} Let the set of active users be $\mathcal{I}=\{1,4,5,6\}$ with demands $D=\{2,3,1,5\}$. By the property of $(\mathbf{P}, \mathbf{B})$ HpPDA, for $\mathcal{I} \subseteq [K], |\mathcal{I}| = K'$, there exists a set $\zeta = \{ S \subseteq \tau \mid |S| = 2 \}$ of size $|\zeta| = 6$ such that 
 $[\mathbf{P}]_{\zeta \times \mathcal{I}} \myeq \mathbf{B}$. Make a new array $\mathbf{\overline{P}}=(\overline{p}_{\tau,k})_{\tau \in \zeta, k \in \mathcal{I}}$ by filling $s \in [S]$ integers in null spaces of the subarray $[\mathbf{P}]_{\zeta \times \mathcal{I}}$ in such a way that $\overline{\mathbf{P}}=\mathbf{B}$. We get,
\begin{align*}
&\begin{matrix}
& 1 & 4 & 5 & 6 
\end{matrix}\\
\overline{\mathbf{P}}=  \begin{matrix}
\{1,4\} \\ \{1,5\} \\ \{1,6\} \\ \{4,5\} \\ \{4,6\} \\ \{5,6\} 
\end{matrix} & \begin{bmatrix}
* & * & 1 & 2 \\
* & 1 & * & 3\\
* & 2 & 3 & * \\
1 & * & * & 4\\
2 & * & 4 & * \\
3 & 4 & * & *
\end{bmatrix}.
\end{align*}
There is a transmission for every $s\in[4]$. Equivalently, we can say that there is a transmission for every $\mathcal{S}\in \mathcal{I}, |\mathcal{S}|=3$. So, the server transmits the following 
\begin{align*}
    X_{\{1,4,5\}}=& V_{\{1,4,5\}} \oplus C_{d_{1},\{4,5\}} \oplus C_{d_{4},\{1,5\}} \oplus C_{d_{4},\{1,5\}}.
    \end{align*}
    \begin{align*}
    X_{\{1,4,6\}}=& V_{\{1,4,6\}} \oplus C_{d_{1},\{4,6\}} \oplus C_{d_{4},\{1,6\}} \oplus C_{d_{4},\{1,6\}}.\\
    X_{\{1,5,6\}}=& V_{\{1,5,6\}} \oplus C_{d_{1},\{5,6\}} \oplus C_{d_{5},\{1,6\}} \oplus C_{d_{6},\{1,5\}}.\\
    X_{\{4,5,6\}}=& V_{\{4,5,6\}} \oplus C_{d_{4},\{5,6\}} \oplus C_{d_{5},\{4,6\}} \oplus C_{d_{6},\{4,5\}}.
    \end{align*}

Consider user~1 and the transmission $X_{\{1,4,5\}}$. User~1 can recover the coded share $C_{d_1,\{4,5\}}$, since it has access to $V_{\{1,4,5\}}$, $C_{d_4,\{1,5\}}$, and $C_{d_5,\{1,4\}}$ from its cache. 
Similarly, user~1 can obtain $C_{d_1,\{4,6\}}$ and $C_{d_1,\{5,6\}}$ from the transmissions $X_{\{1,4,6\}}$ and $X_{\{1,5,6\}}$, respectively. 
In total, user~1 collects $8$ coded shares of its requested file, which are sufficient to reconstruct the file, as a $(15,8)$ MDS code and a $(5,8)$ non-perfect secret sharing scheme are employed. By the nature of $(5,8)$ non-perfect secret sharing, no user knows anything about any file from the caches. Moreover, user~1 does not know the random vector $V_{\{4,5,6\}}$, so it cannot access $X_{\{4,5,6\}}$. Thus, secrecy condition is satisfied.
Similarly, all other users are also able to decode their requested files without compromising the secrecy condition. The rate achieved is $\frac{4}{3}$.

\end{exmp}

\section{Proof of Theorem \ref{thm2}}
\label{scheme2}

In this section, we prove Theorem \ref{thm2}. We present a secretive hotplug coded caching scheme using HpPDAs from $t$-design. 

Consider a $(K,K',F,F',Z,Z',S)$-HpPDA constructed using a $t$-$(v,k,\lambda)$ design. Let us first briefly discuss the construction. Let $(X,\mathcal{A})$ be a $t$-$(v,k,\lambda)$ design with non-repeated blocks. From \cite{Sti}, it follows that, for  $Y \subseteq X$, where $|Y| = s \leq t$, there are exactly $\lambda_s =\frac{\lambda{v-s \choose t-s}}{{k-s \choose t-s}}$ blocks in $\mathcal{A}$ that contain all the points in $Y$. For $ Y \subseteq T \subseteq X$, where $|T| = t, |Y| = i$ with $i \leq t$, there are exactly
$\lambda_i^{t} =\frac{\lambda{v-t \choose k-i}}{{v-t \choose k-t}}$
blocks in $\mathcal{A}$ that contain all the points in $Y$ and none of the points in $T \backslash Y$. An array $\mathbf{P}$ whose columns are indexed by the points in $X$ and rows are indexed by the blocks in $\mathcal{A} $ $(|\mathcal{A}|=b)$ be defined as 
\begin{equation*} \label{arrayP}
\mathbf{P}(\mathcal{A},i)=\begin{cases}
* & \text{if} \ \ i \in \mathcal{A}, \\
null & \text{if} \ \ i \notin \mathcal{A}
\end{cases}.
\end{equation*}
For $0 \leq a_s \leq \lambda_s^{t}$, $1 \leq s \leq t-1,$ consider a set 
$$\mathcal{\mathcal{R}}=\bigcup_{s=1}^{t-1} \left\{ (\mathcal{Y}, i) \mid \mathcal{Y} \in {[t] \choose s}, i \in [a_s] \right\}.$$

Now, another array $B$ whose columns are indexed by the points in $[t]$ and rows are indexed by the elements in $\mathcal{R}$ is defined as 
\begin{equation*} 
\mathbf{B}((\mathcal{Y}, i), j)=\begin{cases}
* & \text{if} \ \ j \in \mathcal{Y}, \\
\left(\mathcal{Y}\cup\{j\},i\right) & \text{if} \ \ j \notin\mathcal{Y}
\end{cases}.
\end{equation*}
The pair $(\mathbf{P},\mathbf{B})$ forms a $(K,K',F,F',Z,Z',S)$-HpPDA, where 
\begin{align*}\label{parameters1}
 \nonumber &K=v, K'=t, F=b, F'=|\mathcal{R}|=\sum_{s=1}^{t-1} a_s {t \choose s}, 
Z=\lambda_1,\\ & Z'=\sum_{s=1}^{t-1} a_s {t-1 \choose s-1}, 
 \text{and} \ S=\sum_{s=1}^{t-1} a_s {t \choose s+1}.  
\end{align*}
The secretive hotplug coded caching scheme operates in two phases as follows.

\textit{Placement Phase}: Each file is first encoded using a $(Z,F'-Z'+Z)$ non-perfect secret sharing scheme \cite{CDN}. Each file is divided into $F'-Z'$ equal-sized parts. So,
\begin{equation}
    W_{i}=\{ W_{i,h}: h \in [F'-Z'] \}, \forall~ i \in [N].
\end{equation}
 Generate $Z$ encryption keys for each $i\in [N]$ independently and uniformly from $\mathcal{F}_{2}^{B/(F'-Z')}$. Let those keys be $\{Y_{i,p} : i \in [N], p \in [Z]\}$. Consider a $(F'-Z'+Z) \times (F'-Z'+Z) $ Cauchy matrix \cite{PlX}, $\mathbf{A}$. Then, $F'-Z'+Z$ shares are generated as in (\ref{eqshares}). Now, consider a $(F, F'-Z'+Z)$ MDS code. Then, $F$ coded shares are generated from $F'-Z'+Z$ shares. Let the coded shares of a file $W_{i}, i \in [N]$ be $\{C_{i,f}: f \in [F]\}$. Now, generate $\sum_{s=1}^{t-2} a_s \binom{K}{s+1}$ random vectors independently and uniformly from $\mathcal{F}_2^{B/(F'-Z')}$. Let those vectors be $\{ V_{(\mathcal{S},i)}:  i \in [a_s], \mathcal{S} \in \binom{[K]}{s+1} \} $ for all $1 \leq s \leq t-2$. Then, the placement is done as follows.
 \begin{multline}
     Z_{k}= \{ C_{i,f} : \mathbf{P}_{f,k}=*, i \in [N], f \in [F]\}~ \cup \\
     \left(\bigcup_{s=1}^{t-2} \bigcup _{i=1}^{a_s} \{V_{(\mathcal{S},i)} : k \in \mathcal{S}, \mathcal{S} \in \binom{[K]}{s+1}\}\right).
 \end{multline}

By the above placement, the size of each cache is
\begin{equation}
    M= \frac{NZ}{F'-Z'}+\frac{1}{F'-Z'} \sum_{s=1}^{t-2} a_s \binom{K-1}{s}.
\end{equation}
 \textit{Delivery Phase}: After the active users reveal their demands, the server starts transmitting. Let $\mathcal{I}=\{i_1, i_2, \ldots, i_{K'}\},$ $1 \leq i_k \leq K, \forall ~k \in [K']$, be the set of active users with demands $\mathbf{D}=(d_{i_1}, d_{i_2}, \ldots, d_{i_{K'}})$. By the property of HpPDA, for set $\mathcal{I}$ there exists a subset $\zeta \subseteq [F], |\zeta|=F'$ such that $[\mathbf{P}]_{\zeta, \mathcal{I}} \myeq \mathbf{B}$. There is a one to one mapping between $[t]$ and $\mathcal{I}$, say $\Phi: [t] \rightarrow \mathcal{I}$. For any subset $\mathcal{U} \in [t]$, define
 \begin{equation*}
     \Phi(\mathcal{U}):= \{ \Phi(u) : u \in \mathcal{U} \}.
 \end{equation*}
 Now,  Make a new array $\mathbf{\overline{P}}=(\overline{p}_{f,k})_{f \in \zeta, k \in I}$ by filling the subarray $[P]_{\zeta \times I}$ with $(\mathcal{Y},i)$, where $\mathcal{Y} \in \binom{[t]}{s+1}$, $1\leq s\leq t-1$, and $1 \leq i \leq a _{s} $ in such a way that $\mathbf{\overline{P}}=\mathbf{B}$. The non-star entries in the matrix $\mathbf{\overline{P}}$ are of the form $(\mathcal{Y},i)$, where $\mathcal{Y} \in \binom{[t]}{s+1} $, $1\leq s\leq t-1$, and $1 \leq i \leq a _{s} $. There is a transmission for every $(\mathcal{Y},i)$ in the matrix $\mathbf{\overline{P}}$. For $s=t-1, 1 \leq i \leq a_s, \mathcal{Y}= [t]$, the transmissions are as follows:
 \begin{equation}
     X_{(\mathcal{Y},i)}=  \bigoplus_{\overline{p}_{f,k} = (\mathcal{Y},i), f \in \zeta, k \in \mathcal{I} } C_{d_k, f}.
 \end{equation}
 For $1\leq s\leq t-2, 1 \leq i \leq a_s, \mathcal{Y} \in \binom{[t]}{s+1}$, the transmissions are as follows:
 \begin{equation}
     X_{(\mathcal{Y},i)}= V_{(\Phi(\mathcal{Y},i)} \oplus \bigoplus_{\overline{p}_{f,k} = (\mathcal{Y}),i), f \in \zeta, k \in \mathcal{I} } C_{d_k, f}.
 \end{equation}

 Since, there is a transmission corresponding to each non-star entry, the rate achieved is $R=\frac{S}{F'-Z'}$. 
 
\textit{Correctness}: Every active user should be able to get the file it demanded. Consider a user $k\in \mathcal{I}$ and a transmission $X_{(\mathcal{Y},i)}$ such that $k\in \Phi(\mathcal{Y})$. The user $k$ has access to $V_{(\Phi(\mathcal{Y}),i)}$ because $k \in \Phi(\mathcal{Y})$. The transmissions corresponding to $s=t-1$ are not protected by any random vectors. So, all users can access these transmissions. Since $\mathbf{\overline{P}}$ is a PDA, each non-star entry in the $i_k$-th column will provide user $i_k$ a coded share of the desired file $W_{d_{i_k}.}$ There are a total of $Z'$ stars and $ F'-Z'$ different non-star entries in a column. As there is a transmission for every non-star entry, each active user gets $F'-Z'$ coded shares of the desired file from transmissions. Each user already has $Z$ coded shares of each file in its cache. So, each user gets $F'-Z'+Z$ coded shares of its desired file, from which it can construct $F'-Z'+Z$ shares. By the property of the $(Z, F'-Z'+Z)$ non-perfect secret sharing scheme, the user can reconstruct the entire file. 

\textit{Secrecy}: Each file is encoded with a $(Z, F'-Z'+Z)$ non-perfect secret sharing scheme which ensures that having $Z$ shares doesn't reveal anything about the file. In the placement phase, the number of coded shares of each file that are stored in each cache is $Z$. So, users know nothing about any file from their cache content alone. The transmissions corresponding to $s=t-1$ are required for all users. Therefore, they are not protected by any random vectors. Since all other transmissions are encoded using random vectors, no user can access a transmission that is not intended for them.
 Thus, any user knows nothing about the file other than its demanded file. This completes the proof of Theorem \ref{thm2}.\hfill $\blacksquare$

\begin{exmp}
    Consider a $(7,2, 7)$ hotplug coded caching system. Let us consider the  $(7,2,7,2,3,1,1)$-HpPDA constructed from a $2$-$(7,3,1)$ design. Choose $s=1,a_1=1$.

\begin{align*}
\mathbf{B} &= 
\begin{array}{c}
(1,1) \\ (2,1)
\end{array}
\left[
\begin{array}{cc}
* & (12,1) \\
(12,1) & *
\end{array}
\right], \\[2ex]
\mathbf{P} &= 
\begin{array}{c}
1 \\ 2 \\ 3 \\ 4 \\ 5 \\ 6 \\ 7
\end{array}
\left[
\begin{array}{ccccccc}
* & * &   &   &   &   & * \\
* &   &   & * & * &   &   \\
* &   & * &   &   & * &   \\
  &   &   & * &   & * & * \\
  & * &   &   & * & * &   \\
  &   & * &   & * &   & * \\
  & * & * & * &   &   &   
\end{array}
\right].
\end{align*}
\textit{Placement Phase}: Each file is encoded using a $(3,4)$ non-perfect secret sharing scheme\cite{CDN}. Then, $7$ coded shares are generated from $4$ shares using a $(7,4)$ MDS code. Let the coded shares of a file $W_{i}, i \in [7]$ be $\{ C_{i,1},C_{i,2},\ldots, C_{i,7} \}$. The placement is done as follows:
\begin{align*}
Z_{1}&= \{ C_{i,1}, C_{i,2}, C_{i,3} : i \in [7] \}, \\
Z_{2}&= \{ C_{i,1}, C_{i,5}, C_{i,7} : i \in [7] \}, \\
Z_{3}&= \{ C_{i,3}, C_{i,6}, C_{i,7} : i \in [7] \}, \\
Z_{4}&= \{ C_{i,2}, C_{i,4}, C_{i,7} : i \in [7] \},
\end{align*}
\begin{align*}
Z_{5}&= \{ C_{i,2}, C_{i,5}, C_{i,6} : i \in [7] \}, \\
Z_{6}&= \{ C_{i,3}, C_{i,4}, C_{i,5} : i \in [7] \},\\
Z_{7}&= \{ C_{i,1}, C_{i,4}, C_{i,6} : i \in [7] \}.
\end{align*}
By the above placement, the cache memory size is $21$.\\
\textit{Delivery Phase}: Let the active users set be $\mathcal{I}=\{2,5\}$. Now, the array $\mathbf{\overline{P}}$ is as follows
\[
\mathbf{\overline P} =\begin{blockarray}{ccc}
& 2 &   5\\
\begin{block}{c[cc]}
1 &* & (12,1) \\
2 & (12,1) & * \\
\end{block}
\end{blockarray}.
 \]
Accordingly the transmission made by the server is $X= C_{d_{2},2}\oplus C_{d_{5},1} $. From the transmission, user $2$ gets $C_{d_{2},2}$ and user $5$ gets $C_{d_{5},1}$. Now, they have $4$ coded shares, which are enough to get their demanded file. As the files are encoded using a $(3,4)$ non-perfect secret sharing scheme, users know nothing about the files from their cache alone. From server transmissions, each user gets a single coded share of their demanded file. There is only one transmission that is useful to both users. So, there is no need to encrypt the transmission with a key. Thus, users know nothing about the files other than the demanded file. So, secrecy is satisfied. The rate achieved is $1$. 

\end{exmp}
\begin{exmp}
Consider a $(8,3, 8)$ hotplug coded caching system. The following HpPDA $(\mathbf{P},\mathbf{B})$ is a $(8,3,14,9,7,5,5)$-HpPDA that corresponds to a $(8,3, 8)$ hotplug coded caching system. We choose $a_1=1,a_2=2$.

\[
\mathbf{P}=\begin{blockarray}{ccccccccc}
& 1 & 2 & 3 & 4 & 5 & 6 & 7 & 8 \\
\begin{block}{c[cccccccc ]}
1 &* & * &  &  & * & * & & \\
 2 & &  & * & * &  &  & * & *\\
 3 & & * &  & * &  & * &  & *\\
4 & * &  & * &  & * &  & * & \\
5 & *&  &  & * & * &  &  & *\\
6 & & * & * &  &  &*  & * & \\
7 & *& * & * & * &  &  &  & \\
 8 & &  &  &  & * & * & * & *\\
9 & *& * &  &  &  &  & * & *\\
10 & &  & * & * & * &*  &  & \\
11 & *&  &*  &  &  &*  &  & *\\
12 & & * &  &*  & * &  & * & \\
13 & *&  &  & * &  & * & * & \\
14 & & * & * &  & * &  &  & *\\
\end{block}
\end{blockarray}, 
\]
\textit{Placement Phase}: Each file is divided into $4$ parts and encoded using a $(7,11)$ non-perfect secret sharing scheme\cite{CDN}. Now, using the $11$ shares $14$ coded shares are generated using a $(14,11)$ MDS code. Let the $14$ coded shares of a file $W_{n}, n \in [8]$ be $\{C_{n,1}, C_{n,2}, \ldots, C_{n,14}\}$.
\[
\mathbf{B}=\begin{blockarray}{cccc}
& 1 & 2 & 3  \\
\begin{block}{c[ccc ]}
(12, 1) &* & * &  (123, 1) \\
 (13, 1) & * & (123, 1) & *\\
 (23, 1) &(123, 1) & * & *\\
(12, 2) &* & * & (123, 2)  \\
 (13, 2) & * & (123, 2) & *\\
 (23, 2) & (123, 2) & * & *\\
(1, 1) &* & (12, 1) &  (13, 1) \\
 (2, 1) & (12, 1) & * & (23, 1) \\
 (3, 1) &(13, 1) & (23, 1) & *\\
\end{block}
\end{blockarray}
 \]
Now, generate $\sum_{s=1}^{t-2} a_s \binom{K}{s+1}=\binom{8}{2}=16$ random vectors independently and uniformly from $\mathcal{F}_{2}^{B/4}$. Let those vectors be $\{ V_{\mathcal{Y},i} : i \in [a_1], \mathcal{Y} \in \binom{[8]}{2}\}$. The placement is done as follows:
\begin{align*}
    Z_{1}= \{ C_{n,1}, C_{n,4}, C_{n,5}, C_{n,7}, C_{n,9}, C_{n,11}, C_{n,13} : n \in [8]\} ~\cup \\ \{V_{\mathcal{Y},1} : 1 \in \mathcal{Y}, \mathcal{Y} \in \binom{[8]}{2} \}, \\
    %\cup \{ V_{\mathcal{Y},i} : i \in [2], \\
    %1 \in \mathcal{Y}, \mathcal{Y} \in \Omega_3\}, \\
     Z_{2}= \{ C_{n,1}, C_{n,3}, C_{n,6}, C_{n,7}, C_{n,9}, C_{n,12}, C_{n,14} : n \in [8]\} ~\cup \\ \{V_{\mathcal{Y},1} : 2 \in \mathcal{Y}, \mathcal{Y} \in \binom{[8]}{2} \},\\
     %\end{align*}
     %~ \cup \{ V_{\mathcal{Y},i} : i \in [2],\\
   % 2 \in \mathcal{Y}, \mathcal{Y} \in \Omega_3\}, \\
   %\begin{align*}
     Z_{3}= \{ C_{n,2}, C_{n,4}, C_{n,6}, C_{n,7}, C_{n,10}, C_{n,11}, C_{n,14} : n \in [8]\} ~\cup \\ \{V_{\mathcal{Y},1} : 3 \in \mathcal{Y}, \mathcal{Y} \in \binom{[8]}{2} \},\\
     %\end{align*}
     %\cup \{ V_{\mathcal{Y},i} : i \in [2],\\
%3 \in \mathcal{Y}, \mathcal{Y} \in \Omega_3\}, \\
%\begin{align*}
     Z_{4}= \{ C_{n,2}, C_{n,3}, C_{n,5}, C_{n,7}, C_{n,10}, C_{n,12}, C_{n,13} : n \in [8]\} ~\cup \\ \{V_{\mathcal{Y},1} : 4 \in \mathcal{Y}, \mathcal{Y} \in \binom{[8]}{2} \},\\ %\cup \{ V_{\mathcal{Y},i} : i \in [2],\\
    %4 \in \mathcal{Y}, \mathcal{Y} \in \Omega_3\}, \\
     Z_{5}= \{ C_{n,1}, C_{n,4}, C_{n,5}, C_{n,8}, C_{n,10}, C_{n,12}, C_{n,14} : n \in [8]\} ~\cup \\ \{V_{\mathcal{Y},1} : 5 \in \mathcal{Y}, \mathcal{Y} \in \binom{[8]}{2} \},\\  %\cup \{ V_{\mathcal{Y},i} : i \in [2],\\
    %5 \in \mathcal{Y}, \mathcal{Y} \in \Omega_3\}, \\
     Z_{6}= \{ C_{n,1}, C_{n,3}, C_{n,6}, C_{n,8}, C_{n,10}, C_{n,11}, C_{n,13} : n \in [8]\} ~\cup \\ \{V_{\mathcal{Y},1} : 6 \in \mathcal{Y}, \mathcal{Y} \in \binom{[8]}{2} \},\\ %\cup \{ V_{\mathcal{Y},i} : i \in [2],\\
%6 \in \mathcal{Y}, \mathcal{Y} \in \Omega_3\}, \\
     Z_{7}= \{ C_{n,2}, C_{n,4}, C_{n,6}, C_{n,8}, C_{n,9}, C_{n,12}, C_{n,13} : n \in [8]\} ~\cup \\ \{V_{\mathcal{Y},1} : 7 \in \mathcal{Y}, \mathcal{Y} \in \binom{[8]}{2} \},\\  %\cup \{ V_{\mathcal{Y},i} : i \in [2], \\
%7 \in \mathcal{Y}, \mathcal{Y} \in \Omega_3\}, \\
     Z_{8}= \{ C_{n,2}, C_{n,3}, C_{n,5}, C_{n,8}, C_{n,9}, C_{n,11}, C_{n,14} : n \in [8]\} ~\cup \\ \{V_{\mathcal{Y},1} : 8 \in \mathcal{Y}, \mathcal{Y} \in \binom{[8]}{2} \}.  %\cup \{ V_{\mathcal{Y},i} : i \in [2], 
    %8 \in \mathcal{Y}, \mathcal{Y} \in \Omega_3\}. \\
    \end{align*}
    \begin{figure*}[h]
    \centering
    \includegraphics[width=\linewidth]{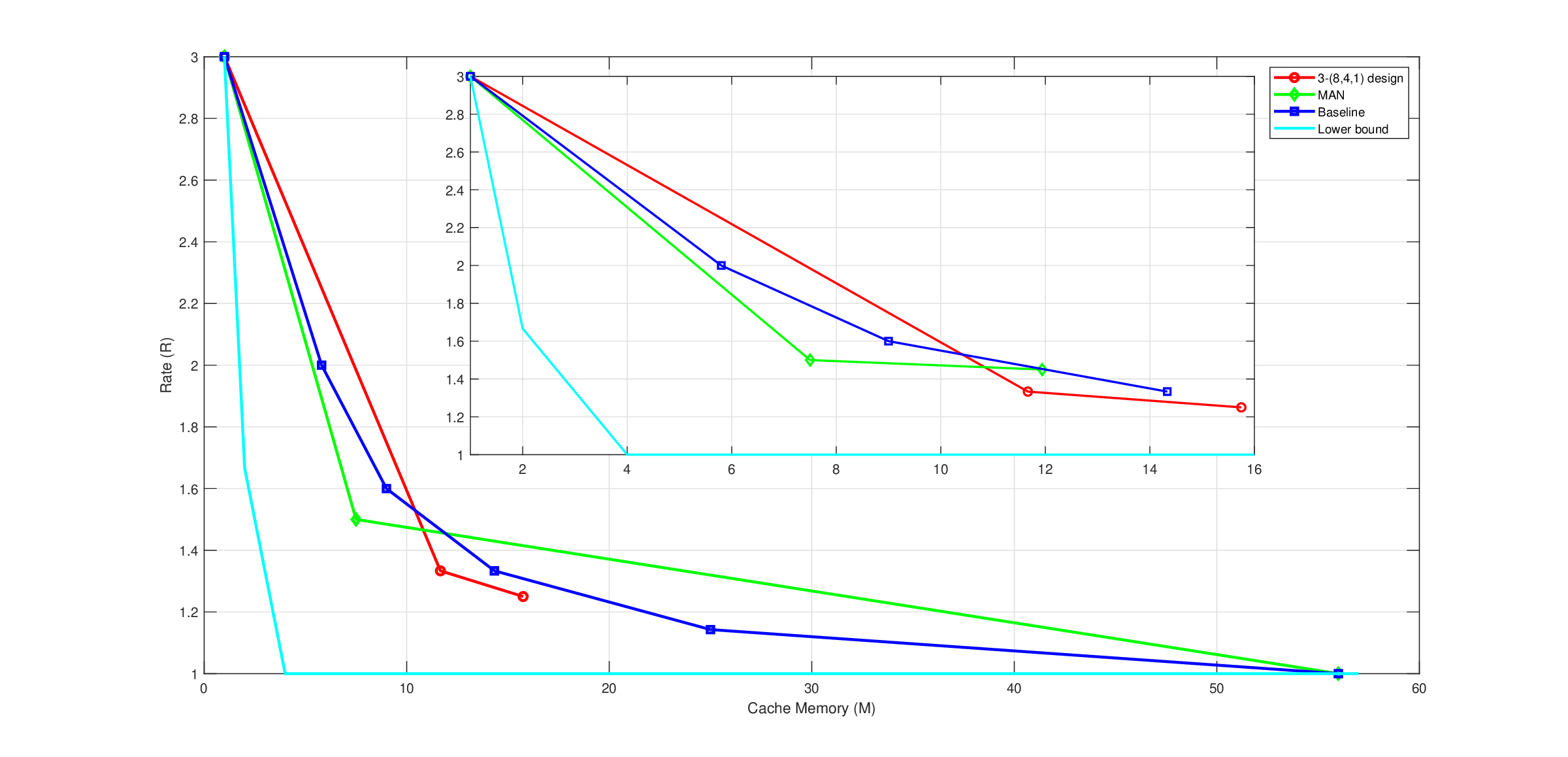}
    \caption{The performance comparison of the proposed schemes with the baseline scheme for a $(8, 3, 8)$ hotplug coded caching system.}
    \label{Fig1}
\end{figure*}

\begin{figure*}[h]
    \centering
    \includegraphics[width=\linewidth]{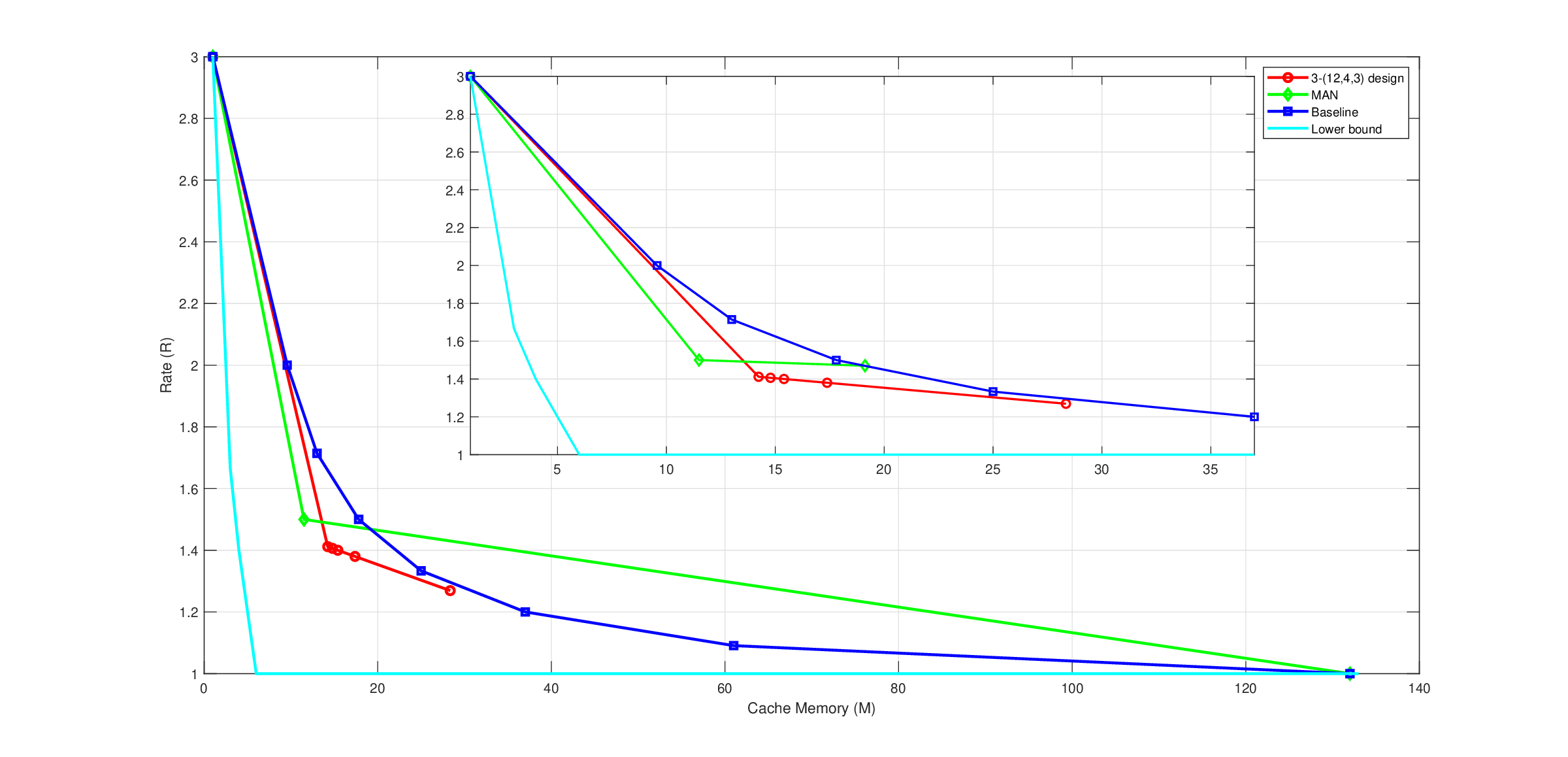}
    \caption{The performance comparison of the proposed schemes with the baseline scheme for a $(12, 3, 12)$ hotplug coded caching system.}
    \label{Fig2}
\end{figure*}
    By the above placement, the size of each cache is $\frac{63}{4}$.\\
\textit{Delivery Phase}: Let the set of active users be $\mathcal{I}= \{2,6,8\}$. Then $\Phi(1)=2, \Phi(2)=6, \Phi(3)=8$. Now, transmissions are done by constructing a new array $\mathbf{\overline{P}}$. By the property of $(\mathbf{P}, \mathbf{B})$ HpPDA, for $\mathcal{I} \subseteq [K], |\mathcal{I}| = K'$, there exists a subset $\zeta \subseteq [F], |\zeta|=F'$ such that 
 $[\mathbf{P}]_{\zeta \times \mathcal{I}} \myeq \mathbf{B}$. Make a new array $\overline{\mathbf{P}}=(\overline{p}_{f,k})_{f \in \zeta, k \in \mathcal{I}}$ by filling non-star positions of the subarray $[\mathbf{P}]_{\zeta \times \mathcal{I}}$ in such a way that $\overline{\mathbf{P}}=\mathbf{B}$. So,
\[
\mathbf{\overline{P}} =\begin{blockarray}{cccc}
& 2 &   6  & 8 \\
\begin{block}{c[ccc ]}
1&* & * & (123,1)\\
9 & *& (123,1) &  *\\
8 &(123,1) & * & *\\
6 &* & * &(123,2) \\
14 &* &(123,2) &* \\
11 &(123,2) & * & *\\
7 & *  &(12,1)  &(13,1) \\
10  &(12,1) & * &(23,1)  \\
 2 &(13,1) &(23,1)  & *\\
\end{block}
\end{blockarray}.
 \]
The server transmissions are given by
\begin{align*}
    X_{(123,1)}&=  C_{d_{2},8} \oplus C_{d_{6},9} \oplus C_{d_{8},1}, \\
    X_{(123,2)}&=  C_{d_{2},11} \oplus C_{d_{6},1} \oplus C_{d_{8},6}, \\
    X_{(12,1)}&= V_{(\Phi(12),1)} \oplus C_{d_{2},10} \oplus C_{d_{6},7} , \\
    X_{(13,1)}&= V_{(\Phi(13),1)} \oplus C_{d_{2},2} \oplus C_{d_{8},7} , \\
    X_{(23,1)}&= V_{(\Phi(23),1)} \oplus C_{d_{6},2} \oplus C_{d_{8},10}.
    \end{align*}

Now, consider user~2 and the transmission $X_{(123,1)}$. The user can recover the coded share $C_{d_{2},8}$, since it already has $C_{d_{6},9}$ and $C_{d_{8},1}$ in its cache. Similarly, user~2 can retrieve $C_{d_{2},11}$ from $X_{(123,2)}$, $C_{d_{2},10}$ from $X_{(12,1)}$, and $C_{d_{2},2}$ from $X_{(13,1)}$. The user has $V_{(\Phi(12),1)} = V_{(26,1)}$ and $V_{(\Phi(13),1)} = V_{(28,1)}$ stored in its cache. Following a similar approach, other active users can also retrieve the coded shares corresponding to their demands. The rate achieved is $\frac{5}{4}$.

Each user stores 7 coded shares in its cache. Due to the properties of the employed $(7,11)$ non-perfect secret sharing scheme, no user gains any information about any file solely from its cached content. Furthermore, since each user only accesses the transmissions that are useful to it, an active user obtains no information about any file it did not request. Therefore, the secrecy condition is satisfied.
  
\end{exmp}

\section{Numerical comparisons} \label{num}

In this section, we numerically compare our scheme with the baseline scheme. One HpPDA corresponds to one memory point for a $(K, K', N)$ hotplug coded caching system. However, corresponding to each $t \in [K']$, we get a different MAN-HpPDA, which corresponds to a different memory point. Therefore, we get a scheme for $K'$ memory points for a $(K, K', N)$ hotplug coded caching system. Similarly, corresponding to one $t$-design, we get multiple HpPDAs, as shown in \cite{RaR2}. Each HpPDA constructed using a $t$-design corresponds to the same hotplug coded caching system with different memory sizes. Using these multiple memory points, we compare the performance of the proposed schemes with the baseline scheme.

In Fig. \ref{Fig1}, we consider a $(8, 3, 8)$ hotplug coded caching system. From a $3$-$(8, 4, 1)$ design, we get two HpPDAs \cite{RaR} with the following parameters:

\begin{enumerate}
    \item a $(8, 3, 14, 9, 7, 5, 5)$-HpPDA,
    \item a $(8, 3, 14, 12, 7, 6, 8)$-HpPDA. 
\end{enumerate}

The performance of the secretive scheme resulting from the HpPDAs from a $3$-$(8, 4, 1)$ design is shown in Fig. \ref{Fig1}, along with the curves for schemes resulting from MAN-HpPDAs, the baseline scheme, and the lower bound given in Lemma \ref{bound}. The secretive hotplug coded caching scheme from MAN-HpPDAs given in Theorem \ref{thm1} is referred to as "MAN" in Fig. \ref{Fig1}. The secretive hotplug coded caching scheme from HpPDAs designed using a $3$-$(8, 4, 1)$ design which is given in Theorem \ref{thm2} is referred to as " $3$-$(8, 4, 1)$ design" in Fig. \ref{Fig1}. It can be observed from Fig. \ref{Fig1} that the Scheme in Theorem \ref{thm1} performs better than the baseline scheme in the lower memory region from $M=1$ to $M=11.9$. The secretive scheme from the $3$-$(8, 4, 1)$ design performs better than the baseline scheme in the memory region from $M=10.4$ to $M=15.7$. The memory ranges are shown in Table \ref{tab1}.

\begin{table}[H] 
\caption{Cache memory ranges where the proposed schemes are better than the baseline scheme for $(8, 3, 8)$ hotplug system. \label{tab1}}
\centering
\begin{tabular}{|c|c|c|c|c|c|} 
 \hline
    & \multicolumn{2}{c|}{From}
            & \multicolumn{2}{c|}{To} \\ 
 \hline
 & $M$ & $R$ & $M$ & $R$ \\
\hline
MAN & $1$ & $3$ & $11.9$ & $1.45$ \\
 $3$-$(8, 4, 1)$ design & $10.4$ & $1.52$ & $15.7$ & $1.25$ \\
\hline
\end{tabular}
\end{table}

Next, we consider a $(12, 3, 12)$ hotplug coded caching system. The performance curves are shown in Fig. \ref{Fig2}. It can be observed from Fig. \ref{Fig2} that the secretive scheme from MAN-HpPDA (Theorem \ref{thm1}), which is referred to as "MAN", performs better than the baseline scheme in the lower memory region from $M=1$ to $M=19.12$. The secretive scheme from $3$-$(12,4,3)$ design (Theorem \ref{thm2}), which is referred to as "$3$-$(12,4,3)$ design", performs better than the baseline scheme in the memory region from $M=1$ to $M=28.3$. The memory ranges are shown in Table \ref{tab2}. In the memory region from $M=1$ to $M=13$, Scheme using MAN-HpPDA (Theorem \ref{thm1}) outperforms Scheme using $3-(12,4,3)$ design (Theorem \ref{thm2}), and from $M=13$ to $M=17.8$, Scheme in Theorem \ref{thm2} outperforms Scheme in Theorem \ref{thm1}.

\begin{table}[H]
\caption{Cache memory ranges where the proposed schemes are better than the baseline scheme for $(12, 3, 12)$ hotplug system. \label{tab2}}
\centering
\begin{tabular}{|c|c|c|c|c|c|} 
 \hline
    & \multicolumn{2}{c|}{From}
            & \multicolumn{2}{c|}{To} \\ 
 \hline
 & $M$ & $R$ & $M$ & $R$ \\
\hline
MAN & $1$ & $3$ & $19.12$ & $1.47$ \\
$3$-$(12, 4, 3)$ design & $1$ & $3$ & $28.3$ & $1.26$ \\
\hline
\end{tabular}
\end{table}
\section{Conclusion} \label{conc}
The secrecy constraint is considered in the hotplug coded caching model. Two different schemes were proposed for two known classes of HpPDAs. The performance comparison was done. The proposed schemes perform better than the baseline scheme in certain memory regions.

\section*{Acknowledgments}
This work was supported partly by the Science and Engineering Research Board (SERB) of the Department of Science and Technology (DST), Government of India, through J.C. Bose National Fellowship to Prof. B. Sundar Rajan.

\end{document}